\documentstyle[preprint,prl,aps]{revtex}
\begin{document}
\draft

\setlength{\topmargin}{-0cm}
\setlength{\headsep}{1.6cm}
\setlength{\evensidemargin}{.7cm}
\setlength{\oddsidemargin}{.7cm}
\setlength{\textheight}{22.5cm}
\setlength{\textwidth}{15.2cm}
\newcommand{\be}{\begin{equation}}
\newcommand{\ee}{\end{equation}}
\newcommand{\ba}{\begin{eqnarray}}
\newcommand{\ea}{\end{eqnarray}}
\newenvironment{technical}{\begin{quotation}\small}{\end{quotation}}

\title{Renormalization of earthquake aftershocks}
\author{Anne Sornette$^{1}$ and Didier Sornette$^{1,2}$}
\address{$^1$ Laboratoire de Physique de la Mati\`{e}re Condens\'{e}e, CNRS UMR 6622\\
Universit\'{e} de Nice-Sophia Antipolis, Parc Valrose, 06108 Nice, France}
\address{$^2$ Department of Earth and Space Sciences and Institute of
Geophysics and Planetary Physics\\ University of California, Los Angeles,
California 90095-1567}

\maketitle

\begin{abstract}

Assume that each earthquake can produce a series of aftershock independently of its size
according to its ``local'' Omori's law with exponent $1+\theta$. 
Each aftershock can itself trigger other aftershocks and so on.
The global observable Omori's law is found to have two distinct power law regimes,
the first one with exponent $p_-=1 - \theta$ for time $t< t^* \sim \kappa^{-1/\theta}$, 
where $0<1-\kappa <1$ measures the 
fraction of triggered earthquakes per triggering earthquake, and the second one with exponent
$p_+=1 + \theta$ for larger times. The existence of these two regimes
rationalizes the observation of Kisslinger and Jones [1991] that
the exponent $p$ seems positively correlated to the surface heat flow: a higher heat flow
is a signature of a higher crustal temperature, which leads to larger strain relaxation
by creep, corresponding to fewer events triggered per earthquake, i.e.
to a larger $\kappa$, and thus to a smaller $t^*$, leading to an effective measured exponent 
more heavily weighted toward $p_+>1$.

\end{abstract}


\pagebreak

\section{Introduction}

Aftershocks constitute a very significant percentage of even the large shallow 
earthquakes [{\it Hough and Jones}, 1997] and all the more so for the small earthquakes that 
dominate the catalogs. An aftershock is usually defined as any earthquake
that occurs within a distance equal to the length of the fault that
ruptured during the mainshock and during the span of time that seismicity rate in that
region remains above its pre-mainshock background level. 

This definition can in fact be
imprecise for a variety of reasons\,: 
\begin{enumerate}
\item  the pre-mainshock rate can be
ill-defined due to inherent intermittent behavior [{\it Kagan and Jackson}, 1991; 
{\it Kagan}, 1994] or poorly constrained by sparse data. The intermittency or clustering in 
time is sometimes quantified by fractal tools [{\it Dattatrayam and Kamble}, 1994;
{\it Henderson}, 1994].
\item  Some earthquakes are sometimes triggered hundreds of kilometers away, 
at significantly greater
distance than the mainshock rupture length [{\it Hill et al.}, 1993;
{\it Steeples and Steeples}, 1996]. 
\item  Patterns of spatial and
temporal self-organization between events at large distance and time scales have been observed
[{\it Knopoff et al.}, 1997; {\it Bowman et al.}, 1998] that suggest more complex
interactions than a simple time delay mechanism between mainshock and aftershocks, and has been
ascribed to a kind of self-organized critical behavior [{\it Sornette and Sornette}, 1989;
{\it Bak and Tang}, 1989; {\it Grasso and Sornette}, 1998];
\item In some models
of self-organized spatio-temporal earthquake behavior
[{\it Huang et al.}, 1998; {\it Hainz et al.}, 1998], the same physical mechanism is found to 
produce both the Gutenberg-Richter frequency-magnitude distribution and the foreshock and 
aftershock Omori's laws.
\end{enumerate}

 The identification of an aftershock becomes feasible, by definition,
only after an event has been recognized as a main shock. This main shock can itself be
an aftershock of a larger event that preceded it. The time-delayed interactions between
mainshock and aftershocks, that 
are usually invoked [{\it Scholz}, 1990; {\it Das and Scholz}, 1981; {\it Shaw}, 1993],
have no reason not to be at work between two events when the second one
is the large.  Kagan and Knopoff [1978] have even asserted that both
foreshocks and aftershocks are a manifestation of essentially the same process.

These considerations lead us to formulate the following hypothesis 
and explore its self-consistency\,:

\vskip 0.5cm
\noindent {\bf Hypothesis}\,: extending the usual definition of an aftershock to allow
for the fact that the delayed
triggered events can be of arbitrary size and not just smaller, can we
construct a coherent model of aftershocks in which 
each earthquake is the generator of a sequence of delayed triggered events
of arbitrary size, in other words, each earthquake is the ``generalized'' aftershock
of a some preceding event? 
\vskip 0.5cm

To address this question, we explore the following model. 
Each event triggers its own ``local'' Omori's
rate of aftershocks. Each of these aftershocks itself triggers its own ``local''
Omori's law of aftershock, and so on in an infinite branching process.
The relevant question for observations is what is
the resulting global Omori's law? Indeed, the individual ``local'' Omori's rate triggered
by each event is not observable, only the superposition of all events at all levels
of the cascade is usually quantified.
This problem belongs to the general class in which
one specifies the laws at the ``microscopic'' scale, here the local Omori's law, and 
strives to derive the resulting ``macroscopic'' laws. In this paper, we do not address
the mechanism(s) at the origin of the ``local'' Omori's law but just take it for granted.
We rather address the collective behavior, i.e. how the ``local'' law is ``renormalized''
[{\it Wilson}, 1979] into a macroscopic observable Omori's law.

\section{Self-consistent Omori's law}

To formulate the question in formal terms, we follow Kagan and Knopoff (1981)
and assume that the probability that one event occurring at time zero
gives birth to another in the time interval between
$t$ and $t+dt$ is
\be
\phi(t) dt = (1-\kappa) ~\theta ~t_0^{\theta}~ {1 \over t^{1+\theta}} ~ H(t-t_0) ,
\label{first}
\ee
where $H(t)$ is the Heaviside function: $H(t-t_0) = 0$ for $t<t_0$ and
$1$ otherwise, which means that the triggering of seismic activity starts 
a short delay $t_0$ after the shock. This minimum waiting time is necessary for the
total number of events to be finite, otherwise $\phi(t)$ would not be
normalized due to the divergence at $t \to 0$.
Physically, it represents the simplest way to account for a progressive
build up of activity after the main shock.

This expression is exactly Omori's law for the rate of aftershocks
following a main shock, albeit with the modification that we do not specify that
aftershocks have to be smaller. The exponent $1+ \theta$ of the ``local'' Omori's law 
has not reason a priori to be the same as the one measured macroscopically which is usually
found between $0.8$ and $1.2$ with an often quoted value $1$. This is in fact the question
we address\,: assuming the form (\ref{first}) for the ``local'' Omori's law, is the global
Omori's law still a power law and, if yes, how does its exponent depend on $\theta$?

The integral of $\phi(t)$ over time is the average number of earthquakes created 
by each event and is equal to
\be
\int_0^{+\infty} \phi(t) ~dt = 1-\kappa .
\label{second}
\ee
If $\kappa < 0$, more than one earthquake is triggered per earthquake. This regime corresponds
to the super-critical regime of branching processes [{\it Harris}, 1963] in which the total
number of events grows exponentially with time. If $\kappa > 0$, there is less than one
earthquake triggered per earthquake. This is the sub-critical regime in which
the number of events following the first main shock decays eventually to zero. The critical
case $\kappa = 0$ is at the borderline between the two regimes.
In this case, there is exactly one earthquake on average triggered per
earthquake and the process is exactly at the critical point between death on
the long run and exponential proliferation. In the sequel, we take $\kappa >
0$ as the physically relevant regime where the aftershock activity following 
a main shock eventually decays at long times.

We analyze the case where there is an
origin of time $t=0$ at which we start recording the rate of earthquakes, assuming that
the largest earthquake of all has just occurred at $t=0$ and somehow reset the clock. In the
following calculation, we will forget about the effect of events preceding the one at $t=0$ 
and count aftershocks that are created only by this main shock. 

Let us call $N(t) dt$ the number of earthquakes between $t$ and $t+dt$. $N(t)$ is the
solution of a self-consistency equation that formalizes mathematically the following 
process\,: the first earthquake may trigger aftershocks; these aftershocks may trigger
their own aftershocks, and so on. The rate of seismicity at a given time $t$ is the
result of this cascade process. The self-consistency equation that sums up this
cascade reads
\be
N(t) = \int_0^t N(\tau) \phi(t-\tau) d\tau ~.
\label{third}
\ee
Its meaning is the following. The rate $N(t)$ at time $t$ is the sum over all induced rates
from all earthquakes that occurred at all previous times with an induced rate per earthquake
occurring at time $t-\tau$ equal to $\phi(t-\tau)$. The essential point is that
the {\it same} function $N$ appears on the l.h.s. and in the integrant in the r.h.s.

The lower bound $0$ in the integral in (\ref{third})
expresses the fact that we look at the time
development of the seismic activity after an origin of time, that can be 
taken for instance as the triggering time at which a big event occurred. Such
an origin of time is important as we are not describing a stationary process,
but rather a rate which on average relaxes (slowly) to zero. 

The problem is then to determine the functional form of $N(t)$, assuming that 
$\phi$ is given by (\ref{first}). The integral equation (\ref{third}) is a
Wiener-Hopf integral equation [{\it Feller}, 1971].
It is well-known [{\it Feller}, 1971; {\it Morse and Feshback}, 1953]
that, if $\phi(\tau)$ decays no slower than an exponential, then
$N(t)$ has an exponential tail $N(t) \sim \exp[-\mu t]$ for large $t$ with 
$\mu$ solution of $\int \phi(x) ~\exp[\mu x]~dx = 1$. 
In the present case, $\phi(\tau)$ decays much slower than an exponential and a
different analysis is called for that we now present.

\section{Solution of the Wiener-Hopf integral equation for power law kernel}

Inserting (\ref{first}) in (\ref{third}) gives
\be
N(t) = (1-\kappa)~\int_0^{t-t_0} N(\tau)  ~ \theta ~t_0^{\theta}~
(t-\tau)^{-(1+\theta)} ~d\tau . 
\label{fourth}
\ee
The upper bound $t-t_0$ comes from the Heaviside function and reflects the
rule (\ref{first}) that the rate $N(t)$ at $t$ is influenced only by events that
occurred at least a time $t_0$ before.

By writing (\ref{fourth}) for time $t+t_0$, we get
\be
N(t+t_0) =  (1-\kappa) \int_0^t N(\tau) ~K_{t_0}(t-\tau) d\tau ~,
\label{fifth}
\ee
where we define the Kernel
\be
K_{t_0}(x) \equiv \theta ~t_0^{\theta} ~(x+t_0)^{-(1+\theta)} ~.
\label{six}
\ee
We recognize that the r.h.s. of (\ref{fifth}) is exactly the convolution of 
$N(t)$ with $K_{t_0}(t)$. Since we are dealing with a non-stationary process (i.e. there
is an origin of time)
and we have a convolution operator, the natural tool is the Laplace transform
${\hat f}(\beta) \equiv \int_0^{+\infty} f(t) e^{-\beta t} dt$. Applying the
Laplace transform to (\ref{six}) yields
\be
\int_0^{+\infty} N(t+t_0) e^{-\beta t} dt = (1- \kappa) {\hat N}(\beta) {\hat
K}_{t_0}(\beta) ~ ,
\label{seven}
\ee
where the r.h.s. has used the convolution theorem. To deal with the l.h.s., we
transform it as
\be 
\int_0^{+\infty} N(t+t_0) e^{-\beta t} dt = 
e^{\beta t_0} \int_{t_0}^{+\infty} N(t) e^{-\beta t} dt =
e^{\beta t_0} \biggl(\int_0^{+\infty} N(t) e^{-\beta t} dt
- \int_0^{t_0} N(t) e^{-\beta t} dt\biggl) .
\label{eight}
\ee
Recall that the first great event is assumed to occur at $t=0$. During the time interval from 
$0$ to $t_0$, no other event occurs. Thus, $\int_0^{t_0} N(t) e^{-\beta t} dt = 
t_0 N(0)$ with $N(0) = 1$. We also make the approximation $e^{\beta t_0} \approx 1$, 
based on the condition 
that we look at earthquake rates at long times $t_0 << t$ which translates into the condition
$\beta << t_0^{-1}$ for the Laplace conjugate variable $\beta$.
We thus get 
\be
\int_0^{+\infty} N(t+t_0) e^{-\beta t} dt \approx
{\hat N}(\beta) - t_0 N(0) ~.
\ee
Reporting into (\ref{seven}) and solving for ${\hat N}(\beta)$ gives the
expression of the Laplace transform of $N(t)$:
\be
{\hat N}(\beta) = {t_0 N(0) \over 1 - (1-\kappa) {\hat K}_{t_O}(\beta)}  .
\label{nine}
\ee
The solution for $N(t)$ is derived by taking the inverse Laplace transform of (\ref{nine}).

In this goal, we shall now make use of a well-known result for the expression of the Laplace
transform of a power law. The important point to notice is that $K(t) \sim
t^{-(1+ \theta)}$ for large $t$. Then, its Laplace transform takes the form
(obtained by integrating by part $l$ times, where $l$ the integer part of
$\theta$ {\it i.e.} $l<\theta<l+1$)
$$
{\hat K}(\beta) = e^{-\beta} \biggl(1 - {\beta \over \theta - 1} + ... + 
{(-1)^l \beta^l \over (\theta - 1)(\theta - 2)...(\theta - l)} \biggl) + 
$$
\be
+{(-1)^l \beta^{\theta} \over (\theta - 1)(\theta - 2)...(\theta - l)}
 \int_{\beta}^{\infty} dx e^{-x} x^{l - \theta} .
\ee
This last integral is equal to
\be 
\beta^{\theta} \int_{\beta}^{\infty} dx e^{-x} x^{l - \theta} =
\Gamma(l+1-\theta) [\beta^{\theta} + \beta^{l+1} \gamma^*(l+1-\theta,\beta)] ,
\ee
where $\Gamma$ is the Gamma function ($\Gamma(n+1)=n!$)
and 
\be
\gamma^*(l+1-\theta,\beta)=
e^{-\beta} \sum_{n=0}^{+\infty} {\beta^n \over \Gamma(l+2-\theta+n)}
\ee
is the incomplete Gamma function. We see that $\hat K(\beta)$ 
presents a regular Taylor expansion in powers of $\beta$ up to the order
$l$, followed by a term of the form $\beta^{\theta}$.
We can thus write
\be
\hat K(\beta) = 1 + r_1 \beta + ..... + r_l \beta^l +
 r_\theta \beta^{\theta} + {\cal O}(\beta^{l+1}) ,
\ee
with $r_1 = -\langle t \rangle, \ r_2={\langle t^2 
\rangle \over 2}, ...$ are the moments of $K$. For small $\beta$, 
we rewrite $\hat K(\beta)$ under the form
\be
\hat K(\beta) = \exp\left[\sum_{k=1}^l d_k \beta^k + d_\theta \beta^{\theta}
\right] ,  
\label{laplaform}
\ee
where the coefficient $d_k$ can be simply expressed in terms of the $r_k$'s.
We recognize that this transformation is similar to that from the moments to the
cumulants. The expression (\ref{laplaform}) generalizes the canonical form
of the characteristic function of the stable L\'evy laws [{\it Gnedenko and Kolmogorov}, 1954] for
arbitrary values of $\theta$, and not solely for $\theta \leq 2$ for which they
are defined. The canonical form is recovered for $\theta \leq 2$ for which the
coefficient $d_2$ is not defined (the variance does not exist) and the only
analytical term is $\langle t \rangle \beta$ (for $\mu > 1$).

Here, we are interested in the case $0 < \theta < 1$ for which 
\be
{\hat K}_{t_O}(\beta) = 1 - d (\beta t_0)^{\theta} + h.o.t.
\label{ten}
\ee
where  $h.o.t.$ stands for higher order terms. Reporting in (\ref{nine})
this gives
\be
{\hat N}(\beta) = {t_0 N(0) \over 1 - (1-\kappa) [1 - d (\beta t_0)^{\theta}]}
=  {t_0 N(0) \over \kappa + d (1-\kappa) (\beta t_0)^{\theta}} .
\label{el}
\ee

Two cases must be distinguished.

$\bullet$ $(\beta t_0)^{\theta} << \kappa$ corresponds to $t >> t^* \equiv
{t_0 \over \kappa^{1 \over \theta}}$. In this case, we can expand 
${t_0 N(0) \over \kappa + d (1-\kappa) (\beta t_0)^{\theta}}$, which leads
to 
\be
{\hat N}(\beta) = {t_0 N(0) \over \kappa} [1 - d {1-\kappa \over \kappa} (\beta
t_0)^{\theta}].
\label{fgh}
\ee
We recognize the Laplace transform (\ref{laplaform}) of a power law of exponent
$\theta$, {\it i.e.} 
\be
N(t) \sim t^{-(1+ \theta)}  ~~~~~ \mbox{  for $t>>t^*$  .}
\label{eaz}
\ee

$\bullet$ For $t < t^*$, $d (1-\kappa) (\beta t_0)^{\theta} > \kappa$ and 
(\ref{el}) can be written with a good approximation as
\be
{\hat N}(\beta) = {t_0 N(0) \over d (1-\kappa) (\beta t_0)^{\theta}} \sim 
(\beta t_0)^{-\theta} . 
\label{elkkk}
\ee
Denoting $\Gamma(z) \equiv
\int_0^{+\infty} dt~e^{-t} ~t^{z-1}$, we see that 
$\int_0^{+\infty} dt~e^{-\beta t}~ t^{z-1} = \Gamma(z) \beta^{-z}$. Comparing
with (\ref{elkkk}), we thus get
\be
N(t) \sim t^{-(1 - \theta)}  ~~~~~ \mbox{  for $t<t^*$  .}
\label{eadfgz}
\ee

\section{Discussion}

Treating all aftershocks on the same footing by assuming that each may trigger new
aftershocks with the same ``local'' rate gives a global Omori's law $N(t)$ described by
{\it two} power laws with exponents $p_-=1 - \theta$ for $t<t^*$ and 
 $p_+=1 + \theta$ for $t>t^*$, where the characteristic time $t^* = t_0/\kappa^{1/\theta}$.
 
 \vskip 0.5cm
\noindent $\bullet$ Our results shows that the simple and parsimonious model in which
 all earthquakes are put on the same footing, i.e. are all susceptible to 
 trigger its train of aftershocks is fully consistent with empirical observations. It suggests
that the usual taxonomy identifying aftershocks as special events may not resist
a more physically-based analysis.
 
\vskip 0.5cm
\noindent $\bullet$ The value $\theta \to 0$, corresponding to a local
 Omori's exponent equal to one, is the only value
 that gives a completely self-similar rate, i.e. the same
 power law for all times, and in addition the same decay rate at the ``local''
 and global level. In other words, the $1/t$ law is the only one that solves the self-consistent
 condition that the same law describes all levels of aftershock triggering processes.
 
\vskip 0.5cm
\noindent $\bullet$  In practice, even a large deviation from $\theta = 0$, say, 
 $\theta = 1/2$
may produce a reasonable approximation to the usual $1/t$ Omori's power law, due to
the existence of the long cross-over between the two power laws whose
exponents possess the value $1$ as their exact barycenter. In this picture,
the observed Omori's law is not exactly $1/t$ but a mixture of two power laws 
slowly crossing over from one to the other. This can also provide a mechanism for the
variations of measured Omori's exponent [{\it Kisslinger and Jones}, 1991]. 

\vskip 0.5cm
\noindent $\bullet$  Let us take $\theta = 0.2$ and $\kappa = 10^{-1}$. This
 yields $t^* = 10^{5}~ t_0$ for the characteristic time scale. If $t_0 = 1~s$, then
$t^* \approx 1$ day. Suppose now that $\kappa=0.05$. This
yields $t^* = 3.2~10^{6}~ t_0 \approx 30$ days. Thus, a small variation of 
the fraction $1-\kappa$ of earthquakes triggered per earthquake (from $0.9$ to
$0.95$) leads to a dramatic variation (day to month) of the characteristic time scale $t^*$.
This offers a simple and physically appealing explanation for
the observations of Kisslinger and Jones [1991] that
 Omori's exponent $p$ seems to be positively correlated to the surface heat flow\,: 
 since a higher heat flow
 is probably a signature of a higher crustal temperature, this leads to larger strain relaxation
 by creep. As a consequence, less strain is accomodated by earthquakes and there are thus
 fewer events triggered per earthquake, meaning that $\kappa$ is larger and $t^*$ is
 smaller. Now, the effective exponent $p$ measured over a time interval $t$
 is approximately an average of $p_-$ and $p_+$
 weighted respectively by $t^*-t_0$ and $t-t^*$. If $t^*$ decreases (resp. increases), 
 the effective exponent $p$ increases toward $p_+$ (resp. decreases toward $p_-$).
 Our theory thus views Omori's exponent as a direct gauge of the fraction of strain accomodated
 by earthquakes. This prediction can be tested by future investigations.

\vskip 0.5cm
\noindent $\bullet$ Kagan and Knopoff [1981] have used a similar model 
to the one studied in the present paper
in order to generate synthetic catalogs. 
Their model differs in the definition of the events,
in the introduction of an additional exponential relaxation and in the introduction of 
a threshold for detection. These complications do not allow for an analytic treatment as
performed here but their numerical simulations are compatible with our results, in particular
with respect to the appearance of a long time scale $t^* = t_0/\kappa^{1/\theta}$.

\vskip 1cm
We acknowledge stimulating discussions with Y. Kagan and L. Knopoff.

\vskip 1cm
{\bf References}
\vskip 0.5cm

Bak, P., and C. Tang, 
Earthquakes as a self-organized critical phenomenon, J. Geophys. Res., 94, 15,635-15,637, 1989.

Bowman, D.D., G. Ouillon, C.G. Sammis, A. Sornette and D. Sornette,
An Observational test of the critical earthquake concept, J.Geophys. Res., 103,
24359-24372,1998.

Das, S., and C.H. Scholz, Theory of time-dependent rupture in the Earth, 
J. Geophys. Res., 86, 6039-6051, 1981.

Dattatrayam, R.S. and V.P. Kamble, Temporal clustering of earthquakes - A fractal approach,
Current Science, 67, 107-109, 1994.

Feller, W., An introduction to probability theory and its applications, vol. II, second
edition (John Wiley and sons, New York, 1971).

Gnedenko, B.V., and A.N. Kolmogorov, Limit distributions
for sum of independent random variables (Addison Wesley, Reading MA, 1954).

Grasso, J.R., and D. Sornette,
Testing self-organized criticality by induced seismicity, J. Geophys.Res., 103, 
29965-29987, 1998.

Haintz, S., G. Z\"oller and J. Kurths, Similar power laws for fore- and aftershocks sequences
in a spring-block model for earthquakes, preprint 1998.
      
Henderson, J;, I.G. Main, R.G. Pearce and M. Takeya, Seismicity in North-Eastern Brazil -
Fractal clustering and the evolution of the B value, Geophys. J. Int., 116, 217-226, 1994.

Hill, D.P., P.A. Reasenberg, A. Michael, W.J. Arabaz et al., Seismicity
remotely triggered by the magnitude 7.3 Landers, California, earthquake, Science,
260, N5114, 1617-1623, 1993.

Hough, S.E. and L.M. Jones, Aftershocks\,: Are they earthquakes or afterthoughts?
EOS, Transactions, American Geophysical Union, 78, N45, 505-508 (Nov. 11 1997).

Huang, Y., H. Saleur, C. G. Sammis, D. Sornette,
Precursors, aftershocks, criticality and self-organized criticality, 
Europhysics Letters, 41, 43-48, 1998.

Kagan, Y.Y., Observational evidence for earthquakes as a nonlinear dynamic process,
Physica D, 77, 160-192, 1994.

Kagan, Y.Y. and L. Knopoff, Statistical study of the occurrence of shallow earthquakes, 
Geophys. J. R. Astron. Soc., 55, 67-86, 1978.

Kagan, Y.Y. and L. Knopoff, Stochastic synthesis of earthquake
catalogs, J. Geophys. Res., 86, 2853-2862, 1981.

Kagan, Y.Y. and D.D. Jackson, Long-term earthquake clustering, Geophys. J. Int.,
104, 117-133, 1991.

Kisslinger, C. and L.M. Jones, Properties of aftershocks sequences in Southern
California, J. Geophys. Res., 96, 11947-11958, 1991.

Knopoff, L., T. Levshina, V.I. Keilis-Borok, and C. Mattoni, 
Increased long-range intermediate-magnitude earthquake activity prior 
to strong earthquakes in California, J. Geophys. Res., 101, 5779-5796, 1996.

Morse, P.M., and H. Feshback, Methods of theoretical physics (Mc Graw Hill, New York,
1953)

Scholz, C. H., The mechanics of earthquakes and faulting 
(Cambridge, New York : Cambridge University Press, 1990).

Shaw, B.E., Generalized Omori law for aftershocks and foreshocks from a simple dynamics, 
Geophys. Res. Lett., 20, 907-910, 1993.

Sornette, A., and D.Sornette,
Self-organized criticality and earthquakes, Europhys.Lett. 9, 197-202, 1989.

Steeples, D.W. and D.D. Steeples, Far-Field aftershocks of the 1906 earthquake, 
Bull. Seismol. Soc. Am., 86, 921-924, 1996.
    
Wilson, K.G.,  Problems in physics with many scales of
length, Scientific American, 241, August, 158-179, 1979.

\end{document}